# Super-resolution Live-cell Fluorescence Lifetime Imaging

Raphaël Marchand*[1,2,3], Henning Ortkrass*[4], Daniel Aziz[1,2], Franz Pfanner [1,2,5], Eman Abbas[6], Silvio O. Rizzoli[6], Wolfgang Hübner[4], Adam Bowman[7], Thomas Huser[4], and Thomas Juffmann[1,2]

1    University of Vienna, Faculty of Physics, VCQ, A-1090 Vienna, Austria

2    University of Vienna, Max Perutz Laboratories, Department of Structural and Computational Biology, A-1030 Vienna, Austria

3    Université Sorbonne Paris Nord and Université Paris Cité, INSERM, LVTS, F-75018 Paris, France

4    Biomolecular Photonics Research Group, Faculty of Physics, Bielefeld University, Bielefeld, Germany

5    California Institute of Technology, Pasadena, USA

6    Department of Neuro- and Sensory Physiology, University Medical Center Göttingen, Göttingen, Germany

7    Salk Institute for Biological Studies, La Jolla, USA

* These authors contributed equally to this work

**Abstract:**

Super-resolution Structured Illumination Microscopy (SR-SIM) enables fluorescence microscopy beyond the diffraction limit at high frame rates. Compared to other super-resolution microscopy techniques, the low photon fluence used in SR-SIM makes it readily compatible with live-cell imaging. Here, we combine SR-SIM with electro-optic fluorescence lifetime imaging (EOFLIM), adding the capability of monitoring physicochemical parameters with 156 nm spatial resolution at high frame rate for live-cell imaging. We demonstrate that our new SIMFLIM technique enables super-resolved multiplexed imaging of spectrally overlapping fluorophores, environmental sensing, and live-cell imaging.

The ability to resolve dynamics on sub-cellular scales has revolutionized our understanding of biology and medicine. Fluorescence microscopy provides the sensitivity and specificity required for obtaining spatio-temporal information on internal cellular structure and allows us to determine the distribution of various biochemical compounds. Super-Resolution optical microscopy techniques, such as Stimulated Emission Depletion (STED) microscopy[1], Single Molecule Localization Microscopy (SMLM)[2–4], or MINFLUX[5] have pushed the achievable resolution to the nanometer scale; however, often at the expense of temporal resolution[6]. For live-cell imaging applications, Super-Resolution Structured Illumination Microscopy (SR-SIM)[7,8] presents a good compromise between spatial and temporal resolution: It enables a two-fold improvement in spatial resolution over the diffraction limit, and can operate at imaging rates up to 100 Hz[9,10], with typical illumination intensities that are low compared to other super-resolution techniques. These features make SR-SIM a powerful method to study a wide range of phenomena in cellular and molecular biology, such as resolving chromatin organization[11,12], nuclear pore complexes[13], liver endothelial cell fenestrations[14,15] and many more.

Fluorescence Lifetime Imaging (FLIM)[16–18] has been demonstrated to provide additional biosensing and environmental sensing capabilities[19]. Compared to intensity-based measurements, FLIM is independent of the fluorophore concentration, fluorophore bleaching, excitation and detection efficiencies. These can be sample- (and time-) dependent due to the scattering and absorption in moving samples, which makes it challenging to calibrate intensity-based measurements. FLIM overcomes these limitations, since the fluorescence lifetime (referred to as lifetime) is an intrinsic property of the fluorophore. FLIM has been widely used for monitoring physico-chemical parameters of cells and tissues, as well as to probe biomolecular interactions, with applications in various fields of science and technology, from materials science to biology, medicine, artwork analysis, and forensics[19–21].

Combining super-resolution live-cell imaging techniques with FLIM provides a new avenue to imaging molecular dynamics on cellular and even sub-organelle spatial scales, enabling an entirely new range of studies in cell biology. Super Resolution FLIM (SR-FLIM) has been demonstrated by combining FLIM with STED microscopy[22–24], confocal SMLM[25], image scanning microscopy[26], a combination of confocal SMLM and image scanning microscopy[27], and based on generalized stepwise optical saturation[28]. However, such scanning approaches limit the achievable spatio-temporal bandwidth product, i.e. the number of resolvable spots within the field of view (FOV) times the number of recorded frames per unit of time, which greatly limits the extension of SR-FLIM to live-cell imaging.

SR-SIM, on the other hand, could enable fast super-resolved FLIM if combined with a sensitive widefield FLIM detector. This potential was partially realized in[29–31] where (SR-)SIM was combined with FLIM. However, the FLIM channel was never super-resolved, and widefield detection was realized using a gated intensified imager, which suffers from low quantum efficiency (QE), and loss of non-gated photons. Other potential wide-field FLIM detectors, such as SPAD arrays, or demodulating frequency domain FLIM cameras, lack the number of pixels, or the low pixel noise specifications, respectively, that are required for live-cell SR-SIM.

Recently, wide-field Electro-Optic FLIM (EOFLIM) has been demonstrated to fulfill the specifications required for live-cell imaging[32–34]. EOFLIM disentangles the temporal gating required for measuring lifetimes on the nanosecond scale and detection efficiency and allows for detecting photons before and after the gate, simultaneously. The gating is accomplished using electro-optic crystals (Pockels cells), which offer high transmission and photon throughput. The use of Lithium Tantalate (LiTa) Pockels cell (PC) enables a high numeric aperture (NA), a large field of view (FOV), and driving frequencies up to 80 MHz in a resonant design. The detection can then be done with commercial scientific CMOS cameras offering low noise, high QE, high signal linearity, and a large number of pixels. These advantages enabled demonstrations in single-molecule imaging[32], in resolving the dynamics of action potentials in *Drosophila melanogaster*[33], and in volumetric FLIM imaging using light-sheet microscopy[34].

Here, we combine EOFLIM with SR-SIM. We demonstrate a time-gated EOFLIM detector, offering higher experimental flexibility than resonant designs, which operates at repetition rates of up to 5 MHz, representing a 100x increase compared to previous time-domain designs[32]. Our optimized EOFLIM detector uses a single PC to gate both the vertical and horizontal polarization channels, maximizing photon collection efficiency and thus lifetime measurement precision. Aberrations are minimized to enable diffraction-limited imaging performance[32,33], as required for super-resolution measurements across wide fields of view.

We combine this setup with a custom-built SR-SIM microscope, which was developed to provide compatibility with pulsed laser sources, and which achieves super-resolved frame rates up to 11 Hz. This system is based on an off-axis Michelson interferometer that allows us to generate the interfering excitation beam pairs. Galvanometric scan mirrors are used to select the angle of the interference pattern and the pattern spacing. Our design provides a high imaging speed, compact size and high stability, while also allowing for easy adjustment of the path lengths to allow the use of pulsed laser sources[35,36]. The interference pattern phase is shifted with a novel, cost-efficient phase modulator based on a voice-coil actuated glass window.

We demonstrate SR-FLIM using SIMFLIM at super-resolved frame rates up to 0.1 Hz. We first characterize our setup using fluorescent beads. We then show that FLIM-SIM can be used for multiplexing based on lifetime in a single spectral channel, distinguishing clearly between markers for glial fibrillary acidic protein (GFAP) and mitochondria in rat hippocampal primary neuro-glia co-culture. We then demonstrate live-cell imaging of the dynamics of moving human osteosarcoma (U2OS) cells, and biosensing capabilities, following changes of lifetime most likely due to oxidative stress. Our experiments demonstrate 123 nm spatial resolution in the intensity channel, 156 nm in the lifetime channel, and excellent sectioning capabilities compared to widefield FLIM.

## Results
### Setup

The experimental setup is shown in Fig. 1 (see Fig. S1 for more details), and is composed of two systems, the SR-SIM excitation system, and the EO-FLIM detection system. The SR-SIM excitation setup generates an interference pattern in the sample plane using a pulsed laser source with 532 nm wavelength and 100 ps pulse length (LDH-DFA-530L, PicoQuant GmbH). The laser beam is fiber-coupled to the SIM microscope, collimated and deflected by a two-axis galvanometric scanner (dynAXIS 421, Scanlab GmbH). The beam is focused by a f=100 mm achromatic lens through a 50/50 non-polarizing beam splitter onto a retro-reflector or through a phase modulator onto a mirror in either interferometer arm, respectively. The phase is modulated by a voice-coil tiltable glass plate (XPR-20, Optotune), which allows to modulate the phase at frequencies up to 1 kHz. The optical path lengths of both interferometer arms are similar. The retro-reflector offsets the incoming beam oppositely to the optical axis, providing the second beam for the interference pattern after recombining both interferometer arms. The switching time between different interference pattern angles, given by the galvanometer angle, is 500 µs. Thus, the imaging speed is not limited by the SIM system, but by the maximum frame rate of the camera sensor and the photostability of the sample. For the three different illumination angles, the polarization is rotated for each beam pair to maintain azimuthal polarization for a maximum modulation depth of the excitation pattern. This is performed by pizza-shaped half-wave-plates (Boulder Vision Inc.). The beam pairs are projected by a f=150 mm and a f=100 mm achromatic lens and via a polarization-maintaining dichroic mirror (DM, F48-532PH, AHF Analysentechnik AG) to the back-focal-plane (BFP) of the 60x 1.5 NA objective lens (UPLAPO60XOHR,

EVIDENT Europe GmbH). The field-of-view (FOV) in the sample plane is 55 µm in diameter. The fluorescence light is collected by the objective lens, filtered using the DM and emission filters (EF, center-width (nm): 646-120, 565-20), and imaged onto the input plane (I1) of the EOFLIM system using a custom, Ploessl type tube lens (TL).

The EOFLIM system is also sketched in Fig. 1. The fluorescence light is split into two fields of orthogonal polarization (H and V) using a polarization-splitting system (PSS). The PSS and the following lens introduce a vertical displacement of the two fields, which are imaged in between the two LiTa crystals of the PC (Leysop Ltd., LTA-9-10-AR510-610-DMP). The PC, operated by a high-voltage driver (BME Bergmann, bpp2b5b/PCM-SH50/BCD1e2), gates the fluorescence light at a specific delay time $t_g$ (gating time) after each excitation pulse: At $t_g$, a high voltage $V_\pi$ (2.25, 2.76 kV) is applied to the PC, which leads to a 90 degrees rotation of the polarization of the incoming light. The value of Vpi to apply for full modulation scales linearly with the wavelength $\lambda$ of the fluorescence light. For our PC, we found $V_\pi \approx 3.5\ V/nm \times \lambda$. Another PSS then splits the two fields into ungated (HH and VV) and gated (VH and HV, respectively) channels that are imaged onto four spatially separated regions of a scientific CMOS detector (Prime BSI, Teledyne Photometrics, operating in Combined Gain mode). The intensity ratio between the gated and the (gated + ungated) images can be used to measure the lifetime at every pixel.

Our PC design is based on the principle described in[32] and[33]. The use of a matched pair of LiTa crystals to form a transverse PC enables high repetition rates, fast switching, and imaging at high values of (NA x FOV)[33]: LiTa features low RF dissipation[37], the transverse design yields a low capacitance and enables fast switching. The matched pair reduces on-axis natural birefringence and thermal effects. In the time-domain implementation described here, this enables repetition rates of up to 5MHz for short periods (10 min), and of 1MHz over extended operation periods (1h), currently limited by the thermal dissipation in the driver and limited cooling power of the chiller used in this study. Rise and fall times (10-90) of the gates is 4 ns, and ringing is down to 1% at full modulation.

A typical instrument response function (IRF) of the PC is shown in Fig.1b. Crucial for imaging applications, the LiTa PC provides good gating efficiency over a wide range of incoming angles. Fig.1c shows the isogyre pattern of the PC with an angular acceptance of 80 mrad, which we defined as the angle at which the switching efficiency drops to 95% (indicated by the white circle). Together with the transverse width of the crystal (9 mm), this enables the gating of an etendue as large as G = 1.3 mm²Sr. This is a significant improvement compared to values achieved with KD*P [32], and well compatible with the phase space volumes typically accepted by high-NA objectives used for live-cell microscopy (the NA of 1.5, the index of refraction of the oil immersion of 1.51, and the 55 µm diameter of the FOV, used for this study, leads to an Etendue G = 17 . $10^{-3}$ mm²Sr).

To obtain SR-SIMFLIM images, we first adapt a standard SR-SIM reconstruction procedure[38], which uses the data recorded at 3 different illumination angles and at 5 different phase steps. We first apply this procedure to all four quadrants of the detector, omitting all non-linear filtering steps. Lifetimes are then estimated from the intensity ratio between the gated and the (gated + ungated) images, where overlap with sub-pixel accuracy is first established using a subpixel registration algorithm[39]. More information can be found in the Online Methods section.

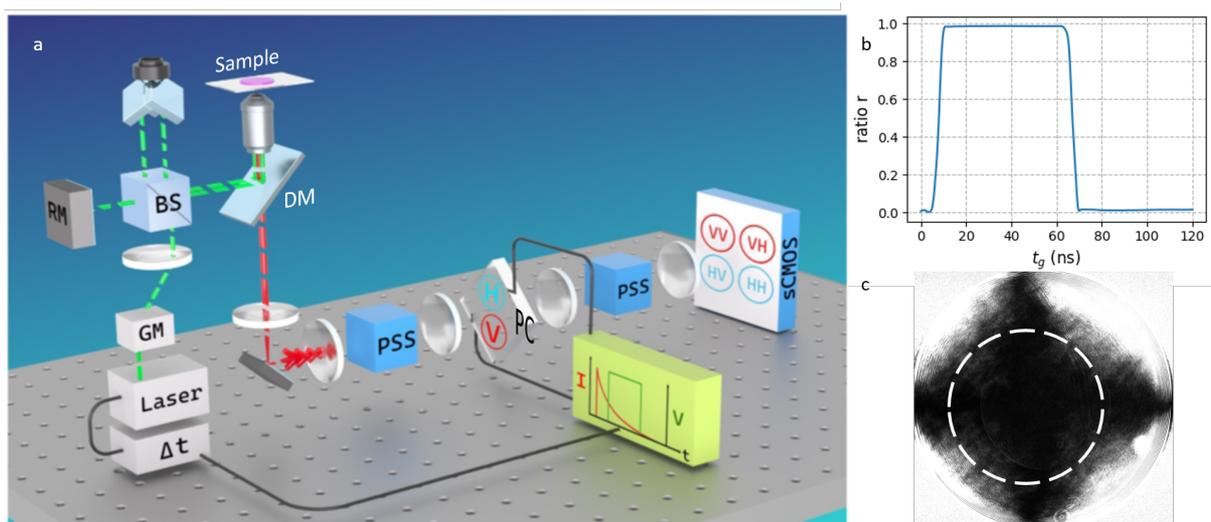

**Fig. 1 | a.** Schematic view of the SIMFLIM microscope system (See Fig. S1 for more details). The laser pulses are fiber coupled to the interferometric SIM system. They are collimated, reflected by a two-axis galvanometric mirror (GM) and split by a 50/50 beam splitter cube (BS). The interferometric paths are reflected by a flat reflective mirror (RM) or a retroreflector that mirrors the beam and displaces it symmetrically to the optical axis. The beam pair is combined again by the beam splitter cube and is projected by a lens telescope via the dichroic mirror (DM) and the objective lens into the sample, where the beams interfere yielding a sinusoidal illumination pattern. The phase shift is performed by modulation of the optical path length of one interferometer path. The fluorescence light collected by the objective is transmitted through the dichroic mirror and emission filter (EF). The first polarization splitting system (PSS) splits the horizontal and vertical polarizations, resulting in two images in the PC marked as H and V. The PC driver (green box) is synchronized with the laser through the delay generator (indicated as $\Delta t$ on the figure). A rectangular voltage pulse (green curve on the driver) is applied to the PC for each excitation pulse. The onset of the rectangular pulse defines the gating time $t_g$ with respect to each excitation pulse, and is controlled by adjusting the delay $\Delta t$ applied to the delay generator. The second PSS splits each polarization channel into a gated and ungated channel, forming a total of four images ij on the sCMOS, where i and j mark the polarization of the light before, and after the PC, respectively. **b.** IRF: average ratio $r(t_g)$ over a region of interest of 100 X 100 pixels (full field of view: 1024 X 1024 pixels) as a function of the gating time $t_g$. Note that the origin of $t_g$ is chosen as the first frame of the IRF, shortly before the onset of the rectangular pulse applied to the PC. **c.** Isogyre pattern obtained by focusing a vertically polarized laser beam into the PC, and imaging the horizontal polarization after the PC. The circle shows the 80 mrad angular acceptance that is gated with an efficiency better than 95%.

**Super-resolved FLIM with SIMFLIM**

As a proof-of-concept, we first demonstrate super-resolved FLIM using SIMFLIM on a sample containing a mix of TetraSpeck (TS) and Fluosphere (FS) fluorescent microspheres (Fig. 2). Fig. 2a shows a wide-field (WF-)FLIM image, where two types of microspheres of different brightness and lifetime can be seen. The brighter (max 30 kcounts/pixel/s) microspheres, with lifetime 3.18 ± 0.07 ns are attributed to the FS beads, while the fainter (max 5 kcounts/pixel/s) microspheres, with a lifetime 3.63 ± 0.14 ns are attributed to the TS microspheres, based on measurements on pure samples (see supplementary Fig. S3).

Fig. 2b shows the corresponding SIMFLIM image, where single TS and FS microspheres which cannot be resolved in the WF-FLIM image can now be resolved (Fig. 1c, 1d, Fig. S4). The spatial resolution of the intensity channel of the SIMFLIM and the WF-FLIM images were estimated to be 141 nm, and 245 nm, respectively, by Fourier ring correlation[40] (Fig. S4).

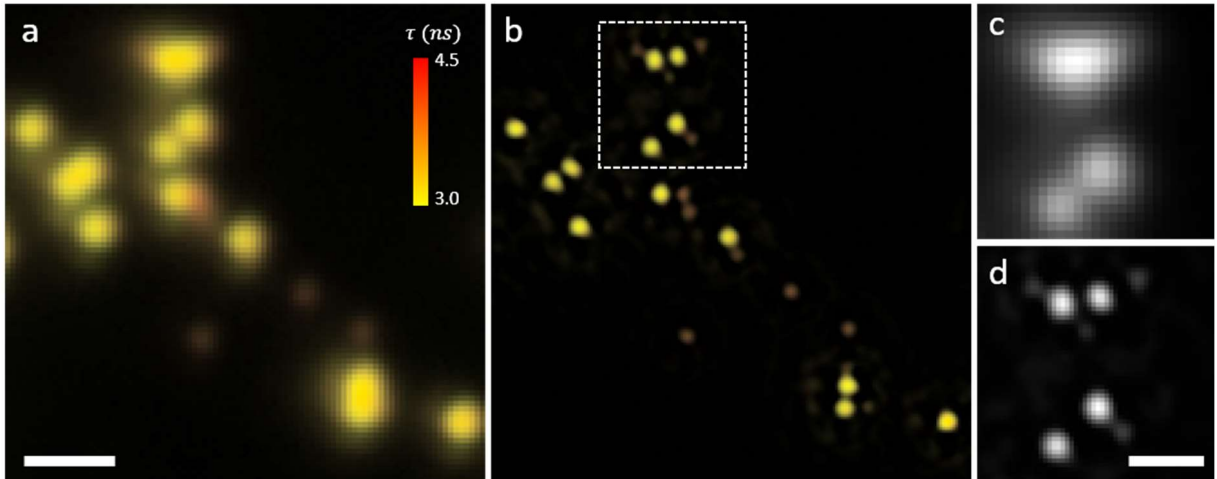

**Fig. 2 | a.** WF-FLIM image of a sample of a mix of FS and TS microspheres. Scale bar: 1 µm. **b.** Corresponding SIMFLIM image. **c, d.** Zoomed views of the intensity channel of a. and b., respectively, corresponding to the white box in b. Scale bar: 500 nm.

**Multiplexing with SIMFLIM**

We then demonstrate multiplexing with SIMFLIM by imaging a sample containing two different fluorophores overlapping spectrally: Fig. 3a shows a SIMFLIM image of a rat hippocampal primary neuro-glia co-culture immunostained for the astrocyte marker GFAP (fluorophore: BDP-TMR) and for the mitochondrial marker TOM20 (fluorophore Cy3). The mitochondria and the GFAP fibers can be separated based on the lifetime (Fig. 3b/c for SIMFLIM and WF-FLIM, respectively), although they are in the same spectral channel and are indistinguishable based on the fluorescence intensity channel only (Fig. 3d/e, respectively). The lifetime of the two fluorophores was found to be 3.7 ns ± 0.15 ns (1 sigma) for BDP-TMR and 2.7 ± 0.7 ns for Cy3 based on a fit of the lifetime histograms (Fig. S5).

The increase of resolution from WF to SR-SIM allows to discriminate details in the SIMFLIM image (Fig. 3b) that are not visible in the WF-FLIM image (Fig. 3c), such as individual GFAP fiber bundles. Fig. 3d and 3e show the corresponding intensity channel. A cross section along the solid white line is shown in Fig. 3f, clearly demonstrating the super-resolution capabilities of the SIMFLIM acquisition.

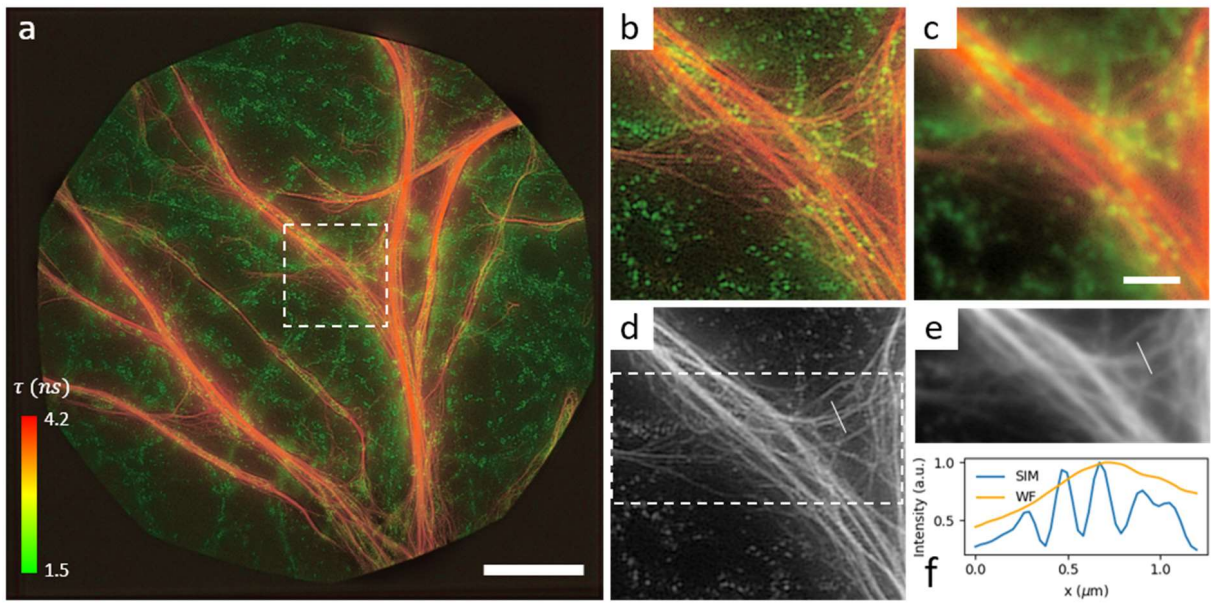

**Fig. 3 | a.** SIMFLIM image of rat hippocampal primary neuro-glia co-culture immunostained for the astrocyte marker GFAP (fluorophore: BDP-TMR) and for the mitochondrial marker TOM20 (fluorophore

Cy3). The two markers can clearly be discerned based on their lifetime. Scale bar: 10 μm. A gamma correction with gamma = 0.4 was applied on the intensity channel (see Online Methods) for better visualization. **b.** Zoomed view of the region of interest (white dashed box) defined in a. **c.** Corresponding WF-FLIM image **d.** Intensity channel corresponding to b. **e.** WF intensity channel corresponding to the region of interest defined in d (white dashed box). Scale bars for b-e: 2 μm. The gamma correction for b-e was done with gamma = 0.5. **f.** Cross-section of normalized intensities (without gamma correction) along the white solid lines in d and e, showing the resolution improvement for SIM.

**SIMFLIM of live-cell dynamics**

We next demonstrate SIMFLIM on live U2OS cells, which requires both fast acquisition and high photon detection efficiency, due to the limited photon budget available for single frames of a dynamic study on such samples. The sequence of images of Fig. 4 shows the dynamics of the membrane and filopodia of a U2OS cell, stained with Vybrant DiI. We take 1 super-resolved frame every 16 seconds, observing the motion of vesicles, the movement and growth of filopodia (red arrow in the inset of Fig. 4) and the endoplasmic reticulum, and a retraction of the cell from certain areas (see inset in Fig. 4), possibly due to phototoxicity. These dynamics can be best appreciated in the supplementary movie SM1, and the corresponding intensity channel alone in supplementary video SM2.

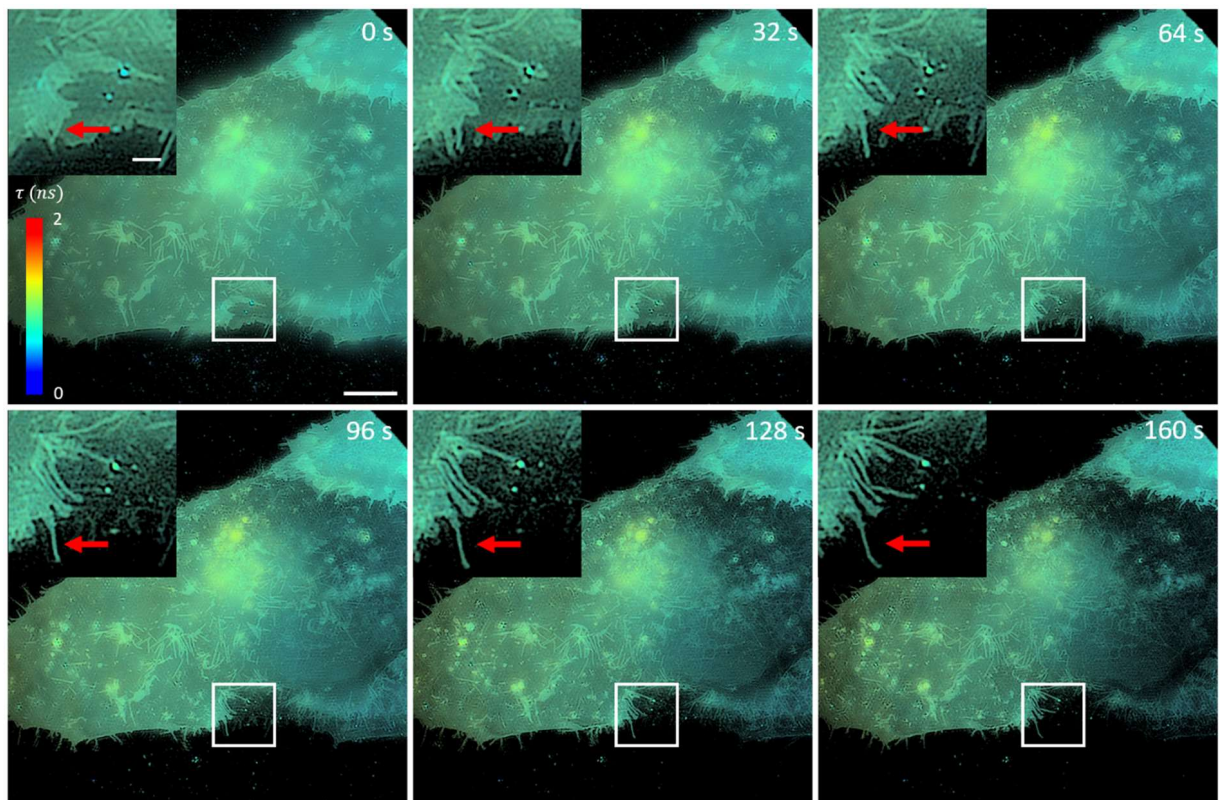

**Fig. 4 |** SIMFLIM image sequence of U2OS cells stained with Vybrant DiI. The gamma of the intensity channel was adjusted to 0.5. Scale bar: 5 μm. The inset shows a zoomed view of the white box, highlighting the retraction of the cell from the corresponding area and the movement and growth of filopodia (see for example the red arrow). Corresponding scale bar: 1 μm. The elapsed time starting from the beginning of the acquisition is shown on the top right corner of each image. The sequence was extracted from the supplementary video SM1.

**SIMFLIM demonstrating super-resolved environmental sensitivity**

Lastly, we present our technique in the context of environmental sensing. Fig. 5 shows a U2OS cell stained with MitoTracker Orange CMTMRos (MTO). We observe both morphological changes and changes of the MTO lifetime as the cell undergoes apoptosis, most likely caused by photo-induced oxidative stress. We observe fluorescently labelled mitochondria in a transition from tubular to globular shape (supplementary movie SM3, SM4), a commonly observed morphological change following photo-induced oxidative stress[41–43]. After transformation into globular shape, the mitochondria further transition into a hollow-looking structure with circular rims and patches attributed to cristae being relocalized to the edge[44–46].

While these morphological changes occur, we also observe a release of fluorophore from the mitochondria to the cytosol and the nucleus, as suggested by the transient increase in fluorescence signal in these areas in the first few frames, before signal fainting due to photobleaching. Relocalization of MTO into the nucleus, and an increase of the MTO lifetime as a consequence of oxidative stress have been reported in[47]. This allows us to measure the lifetime in various cellular environments, as apparent in Fig. 5 and further analyzed in Fig. S6 a. Furthermore, we also observe a general increase of lifetime over time, with the mitochondria showing the most drastic change while they undergo the structural transformation described above, indicating the drastic change of the physicochemical environment of the fluorophore (Fig. S6 b). After transformation of the mitochondria into a hollow-looking round structure, some patches attributed to relocalized cristae (see above) keep a shorter lifetime compared to the surroundings (Fig. 5c, well visible on the edges of mitochondria in Fig. S7 b), indicating a different fluorophore physico-chemical environment.

When comparing the SIMFLIM images to the WF-FLIM images, we again see an improvement in transverse resolution and contrast (Fig. 5 b-c, yellow arrow), translating into super-resolved lifetime measurements (Fig. S7 b, e, f). Using FRC, the spatial resolution in the intensity channel was estimated to be 123 nm, and 156 nm in the corresponding lifetime channel. Importantly, the SIMFLIM data also profits from the sectioning capabilities of SIM[48]. Since out-of-focus fluorescence may have a different lifetime than the structure in focus, sectionning leads to FLIM images with better contrast, spatial resolution, and smaller systematic errors.

While our technique enables fast super-resolution lifetime imaging, our results also show the need for fluorophores of low phototoxicity. This is evident from the lifetime changes observed in Fig. S6 which are likely induced by phototoxicity at an illumination intensity of 25 W/cm². This problem is not specific to our technique, but to any fluorescence-based microscopy scheme that operates at comparable intensities.

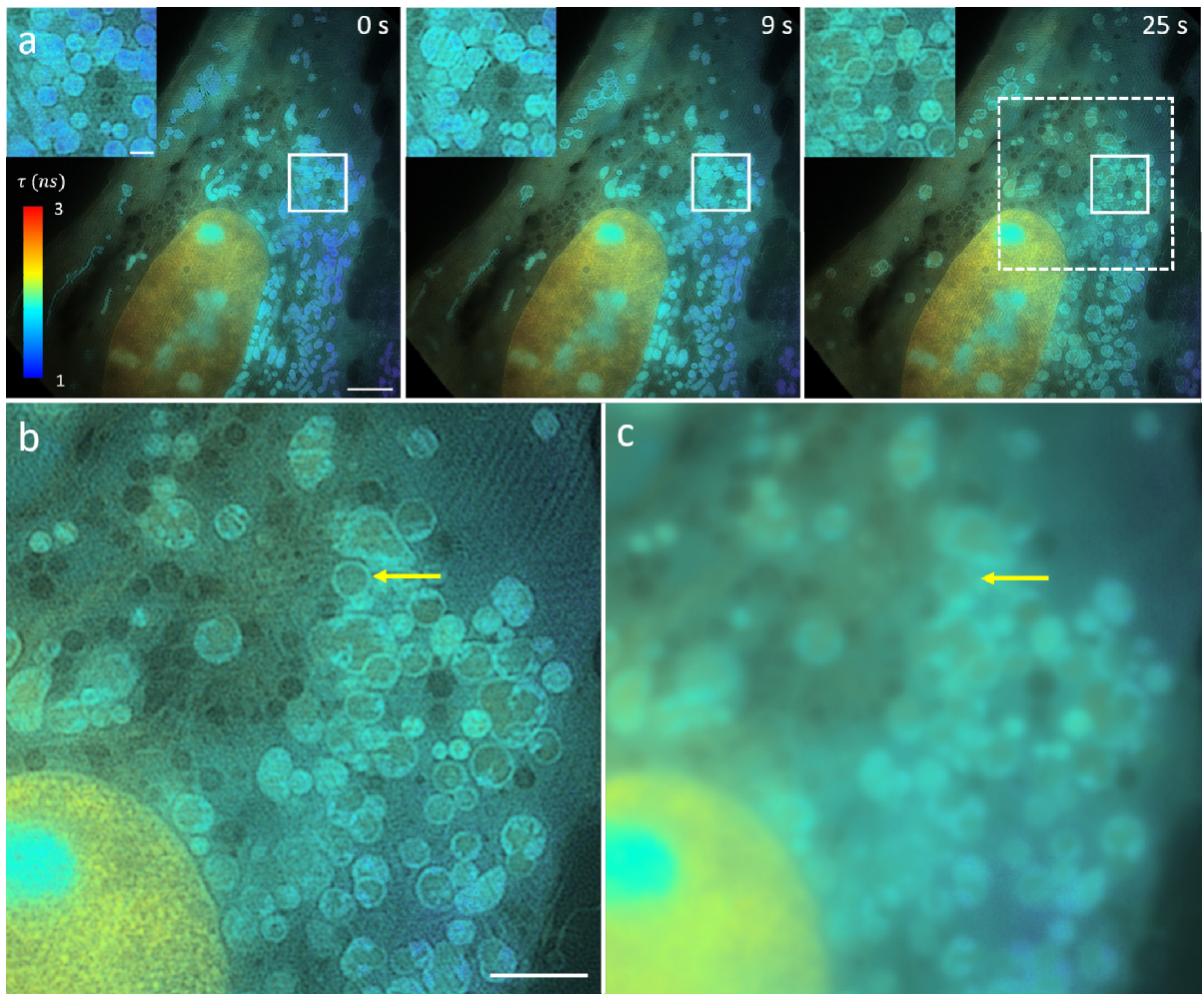

**Fig. 5** | **a.** SIMFLIM images sequence of a U2OS cell stained with MitoTracker Orange CMTMRos (MTO), extracted from supplementary movie SM3. The inset shows a zoomed view of the white solid box, highlighting the transformation of mitochondria into a circular hollow-looking structure. The elapsed time starting from the beginning of the acquisition is shown on the top right corner of each image. Scale bar 5 μm (1 μm in inset). **b.** zoomed view on the white dashed box in a. Scale bar: 3 μm. **c.** corresponding WF-FLIM image. The yellow arrow highlights the resolution and optical sectioning improvement from the WF-FLIM image to the SIMFLIM image, with the edge of the mitochondria appearing blurry in the WF-FLIM image but well defined in the SIMFLIM image.

## Discussion

Our work demonstrates super-resolution lifetime live-cell imaging based on a combination of EO-FLIM and SR-SIM. We show that these two wide-field imaging techniques are fully compatible and that they enable intensity and lifetime measurements at a spatial resolution of about 123 nm and 156 nm respectively, and at framerates of 0.12 Hz, compatible with live-cell imaging.

We achieved this using a LiTa PC, which enables fluorescence gating with MHz repetition rates, and which is compatible with the field of view and numerical aperture typically used in live-cell fluorescence microscopy. Our EOFLIM detector is designed modular and can be added to any wide-field fluorescence microscope with pulsed or intensity-modulated excitation. Our cost-effective EOFLIM design, which makes use of both polarization channels, allows us to collect a high number of photons per pixels (up to $N \approx 10^5$ photons per pixel per super-resolved frame in the data shown here), enabling high lifetime measurement precision ($\propto \frac{1}{\sqrt{N}}$, Fig. S8). Better thermal management will reduce remaining artefacts even further. Achieving such high precision at high framerates, across a mega-pixel image (Fig. 4: 4.2 Mpx, Fig. 5: 1.7 Mpx) is currently impossible in a scanning approach. While resonant frequency-domain EOFLIM[33] is restricted to lifetimes less than 1 over the frequency of the PC modulation, gated EOFLIM allows for more flexibility, as the repetition rate of the gate can be varied from 0 to the MHz range. This potentially enables EOFLIM with fluorophores of long lifetimes (e.g. quantum dots). Gated EOFLIM further allows optimizing the gating function for specific measurements, i.e. for a particular set of lifetimes within a sample. Furthermore, a square gating function provides slightly better sensitivity than a resonantly-driven one.

The SIM-system based on an off-axis Michelson interferometer provides both a compact and cost-efficient design. It is compatible with high power and short coherence-length laser sources. This enables the operation with pulsed lasers, as required for non-linear SIM based on saturation[49] or 2-photon absorption. Such schemes could enhance transverse and axial resolution, and would be ideal for probing endogenous fluorophores in 3D deep inside tissue. Compared to other SIM-systems, the size of the entire SIM-system is very small (34x53 cm²), and the laser power efficiency from the fiber to the objective lens is high (40%). By employing galvanometric scan mirrors for selecting the excitation pattern angle and tiltable glass windows for modulating the pattern phase, the system provides transition times between different excitation patterns below 1 ms, shorter than the readout time of the camera. Due to the achromatic design, the setup potentially also supports multi-color excitation.

The combination of low-aberration EOFLIM with SR-SIM enables super-resolution live-cell imaging. Specifically, we demonstrate lifetime-based multiplexing in a rat hippocampal primary neuro-glia co-culture, the ability to follow dynamics of live U2OS cells, and the environmental sensing under photo-induced oxidative stress. These studies benefit not only from the increased transverse spatial resolution of SIM, but also from its optical sectioning capabilities comparable to those of confocal microscopy [48] leading to better contrast. Our work shows that, using EOFLIM, lifetime information can be gained with high sensitivity, spatial or temporal resolution, opening a new information channel in (super-resolved) live-cell imaging. We expect this capability to greatly improve the spatiotemporal dynamics of intracellular sensing based on novel fluorescent proteins[50,51], and to open new possibilities based on multiplexing. For example, analyzing neuronal phenotypes requires visualizing various elements in parallel[52]. Similarly, approaches related to cell biology require the determination of the positions of numerous organelle markers, from the nucleus, ER and mitochondria to the different endosome types. In such applications, spectral multiplexing is typically limited to 3-4 channels. Fluorescence-lifetime-based multiplexing[53] adds to these capabilities in a multiplicative way and enables efficient multiplexing in live-cell imaging.

## Acknowledgements

We acknowledge useful discussions with Francisco Balzarotti, Lukas Hille, Nele Jarnot, Clara Conrad-Billroth, Philipp Weber, Jörg Enderlein, and Ivana Matoušová Víšová for help with the acquisition of the isogyre patterns. Marcel Müller is acknowledged for software support. We thank the biooptics facility of the Max Perutz Labs for support. We also acknowledge PicoQuant GmbH for the loan of the laser source free of charge. We also acknowledge the Büllesbach family, the Wegscheid family, and flatmates from August-Bebel-Straße, Bielefeld, for their kind accommodation offer free of charge. This project was supported from the ERC Grant 758752 – MICROMOUPE, the ERC PoC Grant 101069260 – EOFLIM and the EIC PATHFINDER Open Programme No 101046928 - DeLIVERy. R.M. acknowledges funding from the European Union's framework program for research and innovation, Horizon 2020 (2014–2020) under the Marie Curie Skłodowska Grant Agreement No. 847548.

## Online Methods

**Samples preparation**

Microsphere sample: The fluorescent beads are a mixture of 200 nm TetraSpeck™ beads (T7280, Thermo Fisher Scientific Inc, nominal diameter 200 nm) and FluoSpheres™ (Invitrogen, nominal diameter 170 nm). Both stock solutions were diluted 10 times with bidistilled water and mixed. 30 µl of the vortexed suspension was dropped on a #1.5 cover glass and dried at room temperature upside down. The cover glass was mounted with a 1:1 mixture of bi-dist. water and glycerol as an embedding medium on a microscope slide and sealed with nail polish. This ensures a refractive index of 1.40 of the medium for low spherical aberrations.

For the recording of the IRF, 25 mg of Allura Red AC (analytical standard, Merck) was added to 1 mL (or to 100 µL when more signal was needed) of bidistilled water, mixed thoroughly and stored at 2°C. 1 µL of this solution was introduced in a small hole (1 mm wide) punched in an adhesive silicon pad (1 mm thick, Grace Bio-labs) and stuck on a cover glass (24 mm X 50 mm). The evaporation was prevented by placing a cover glass (18 mm X 18 mm) on top of the silicon pad. Alternatively, 5 µL of this solution were introduced between a microscopy glass slide and a cover glass (18 mm X 18 mm) and sealed with nail polish.

Fixed neuron culture: The neuron sample is a rat hippocampal primary neuro-glia co-culture containing neurons and astrocytes, cultured for 14 days in vitro, prepared exactly as previously described[54]. The cells are immunostained for the astrocyte marker GFAP with the fluorophore BDP-TMR (Cat#: N3805, conjugation with fluorophore BDP-TMR; NanoTag) and for the mitochondrial marker TOM20 (Cat#: 11802-1-AP; Proteintec) with the fluorophore Cy3 (Cat#: 111-165-144; Diavona), following the standard procedures as previously described[54].

Live-cell samples: U2OS cells used for live imaging (Fig. 4, 5) were labeled with Vybrant DiI and MitoTracker Orange CMTMRos cell labeling solution according to the manufacturer instructions (Thermofisher scientific) to visualize plasma membrane and mitochondria, respectively.

**Data acquisition**

The data acquisition was done according to the scheme described in Fig. S1 and S2. In order to reconstruct one SIM frame, 15 raw frames (exposure time: 500-800 ms) with three different illumination pattern angles and five different phases were acquired with the custom SIM setup. The custom-built SIM microscope allows transition times between the illumination patterns of 1 ms, leaving the system speed-limited by the readout time of the camera.

For the static samples (Fig. 2 and Fig. 3), 2 super-resolved frames (2 x 15 raw frames) were taken at two different $t_g$ values (10.4 ns, 69.6 ns, corresponding respectively to the rising/falling edge of the pulse applied to the PC) and combined as explained in *data processing* below. For the live samples (Fig. 4 and Fig. 5) the movement of the sample prevented from doing similarly, therefore the gating time $t_g$ was fixed to a single value (10.4 ns), and a 15-frames raw stack (one super-resolved frame) was acquired every 16 s (Fig. 4) or 9 s (Fig. 5).

**Data processing**

The raw data of Fig. 2 and 3 consists, for each figure, of 2 stacks of 2048 X 2048 pixels, 15 frames each. One stack is taken with $t_g = 10.4\ ns$, and one with $t_g = 69.6\ ns$. Each stack is processed separately: the quadrants separation converts each stack into 4 (1024 X 1024 pixels, 15 frames) stacks corresponding to the HV, HH, VV, VH quadrants. The SIM processing (see details below) converts each of these stacks into a (2048 X 2048) super-resolved image, with a pixel size corresponding to $(29\ nm)^2$ in the specimen plane. The FLIM processing (see details below) converts these 4 super-resolved images (HV, HH, VV, VH) into one FLIMSIM (2048 X 2048) image, with an intensity and a lifetime channel. The final step combines the SIMFLIM images obtained with the 2 different gating times: the intensity channels are summed, and the lifetime channels are averaged, which results in reduced systematic errors (see Fig. S9).

The raw data of Fig. 4 and 5 consists of one stack for each figure, of 2048 X 2048 pixels, 15 X N frames each, with N the number of frames in the image sequence shown in the movies SM1-4. The processing of each super-resolved frame in these stacks is done as described above, but without the final step of combining data from two different gating times.

*SIM processing:* First, the gated and ungated stacks of each polarization channel are overlapped using an ImageJ plugin[55]. From those overlaid images, the SIM parameters are estimated and applied to the SIM reconstructions of both the gated and ungated channel. Using the same parameters for both channels avoids artefacts in FLIM processing. The low-resolution stack (1024 X 1024 pixels, 15 X N frames) of each quadrant (HV, HH, VV, VH) is then reconstructed into a super-resolved stack (2048 X 2048 pixels, N frames) using the fairSIM ImageJ plugin[38], with and without Wiener filtering. The super-resolved stacks obtained with (resp. without) Wiener filtering are subsequently used to obtain the intensity (resp. lifetime) channel of the SIMFLIM stacks (see FLIM processing below). The WF stacks (N frames) of each quadrant are obtained by summing the 15 raw frames of a SIM (15 X N frames) stack. The filtered WF images are obtained from the fairSIM image reconstruction (one filtered WF frame corresponding to one super-resolved frame).

*FLIM processing: Overlap:* The super-resolved SIM (for SIMFLIM) or WF (for WF-FLIM) stacks of each quadrant (HV, HH, VV, VH) are correlated spatially with sub-pixel accuracy using pyStackReg[39]. Fourier high-pass filtering may be applied beforehand to improve the overlap accuracy. For the live-cell data, the overlap parameters are obtained from the first frame and applied to the whole stack. *Intensity channel:* After background subtraction, the intensity channel of the FLIM stack is obtained by summing the 4 overlapped (HV, HH, VV, VH) stacks. *Ratio calculation:* For each pixel, a ratio, which will be subsequently related to the fluorescence lifetime, is defined from the gated und ungated intensities:

$$r(t_g) = \frac{I_{gated}(t_g)}{I_{gated}(t_g) + I_{ungated}(t_g)} = \frac{I_{HV}(t_g) + I_{VH}(t_g)}{I_{HV}(t_g) + I_{VH}(t_g) + I_{HH}(t_g) + I_{VV}(t_g)}$$

where $I_{ij}(t_g)$ refers to the background-subtracted intensity in the ij quadrant. Using both the H and V polarization channels for the gated (resp. ungated) intensity maximizes the photon collection efficiency and reduces systematic errors (Fig. S10).

*Lookup table (LUT):* From the IRF, a LUT is calculated that relates ratio (at a given gating time $t_g$, chosen specifically for each sample) to lifetime on each pixel (at a resolution of 1024 X 1024 pixels). The calculation convolves a mono-exponential intensity decay $I(t) = I(0)e^{-\frac{t}{\tau}}$ with the pixel-dependent IRF to calculate what $r(t_g)$ would be for a certain lifetime $\tau$. Repeating this for various $\tau$ values gives a table that relates measured intensity ratios to mono-exponential intensity curves characterized by a lifetime $\tau$. For the data represented in this manuscript we calculated the convolution for 20 different lifetime values linearly dispersed between $\tau = 0\ ns$ and $\tau = 10\ ns$. We then used linear interpolation to relate a measured ratio to a certain lifetime.

*Composite FLIM images with lifetime and intensity channels:* In the WF-FLIM and SIMFLIM images in our manuscript, the lifetime and intensity channels are combined in a composite FLIM image, as usually done in FLIM: The lifetime value is represented by a color (see respective color bars) and the intensity channel determines the brightness of the color according to a (gamma-corrected when specified) grayscale map. For Fig. 4, an additional step of photobleaching correction using a Fiji plugging[56] is applied before gamma correction.

# Supplementary Figures

## Contents:

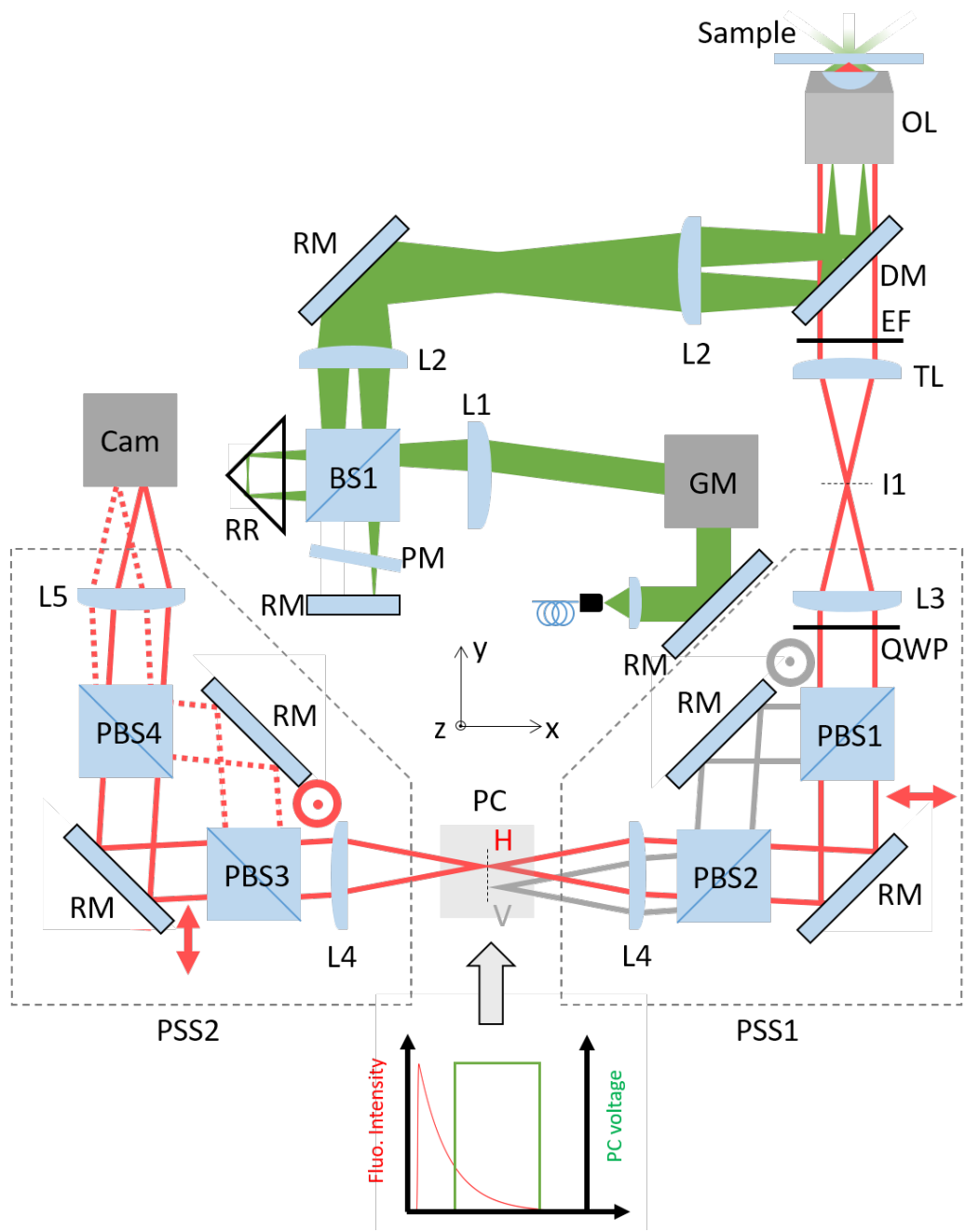

**Fig. S1: Detailed sketch of the experimental setup.** The excitation laser beam is fiber-coupled to the optical setup, collimated and reflected on a two-axis galvanometric scanner (GM), allowing to control the orientation and the period of the interference pattern. The deflected beam is focused by a f=100 mm achromatic lens (L1) through a modified Michelson interferometer: The focused beam is split by a 50/50 beam splitter cube (BS1). While the transmitted path is reflected by a hollow retro-reflector (RR), the reflected path is mirrored by a flat reflective mirror (RM). Both paths are combined by BS1 again, and the resulting beam pair is collimated by a second achromatic lens (L2, f=100 mm). The lateral offset of the beam pair as well as the azimuthal orientation is controlled by the deflection angle of the galvanometric scanner. For the three illumination pattern angles needed for SIM[1], the galvo scanner selects the corresponding deflection angles[2]. The phase of the interference pattern is shifted by modulating the optical path length of the interferometer path with the phase modulator (PM). The beam pair is projected via a lens (L2, f=100 mm) and a polarization maintaining dichroic mirror (DM)

into the objective lens (OL) and the sample. The fluorescence light collected by the objective lens is transmitted through the dichroic filter and an emission filter (EF) and focused into an image plane I1. For better visualization, we only indicate the fluorescence of one particle localized in the center of the FOV. The first polarization splitting system (PSS1), containing relay achromatic lenses (L3: f=300 mm, L4: f=200 mm), polarization beam splitter cubes (PBS1 and PBS2) and reflective mirrors, splits the horizontal (red, ↔) and vertical (grey, ⊙) polarizations, resulting in two demagnified images in the Pockels cell (PC) marked as H and V. The demagnification ensures the two images fit in the crystal. In the drawing, we omit the V channel after the PC for better readability. The PC changes the polarization of the light during the time window within which the voltage Vpi is applied. The second polarization splitting system (PSS2) splits the light into the gated light (now vertically polarized, dashed line) and the ungated light (horizontally polarized, solid line) which are imaged separately on the sCMOS (Cam) with L5 (f=400 mm). Including the originally vertically polarized light, this leads to four separate images on the camera as indicated in Fig. 1 in the main text. While the drawing indicates the image shifts induced by PSS1 and PSS2 in the xy-plane, in the experiment PSS1 shifts along the z-axis instead. In general, the point spread function of the two polarization channels H and V depends both on the polarization of the incoming light and the orientation of the molecular dipoles[3]. For example, the image of fluorescent microspheres was found to be elongated in one direction for one polarization channel, and in the orthogonal direction for the other polarization channel, as previously observed[4]. To mitigate this effect, we insert a quarter waveplate to mix the polarization channels before the splitting by the PBS1, which significantly reduces the ellipticity.

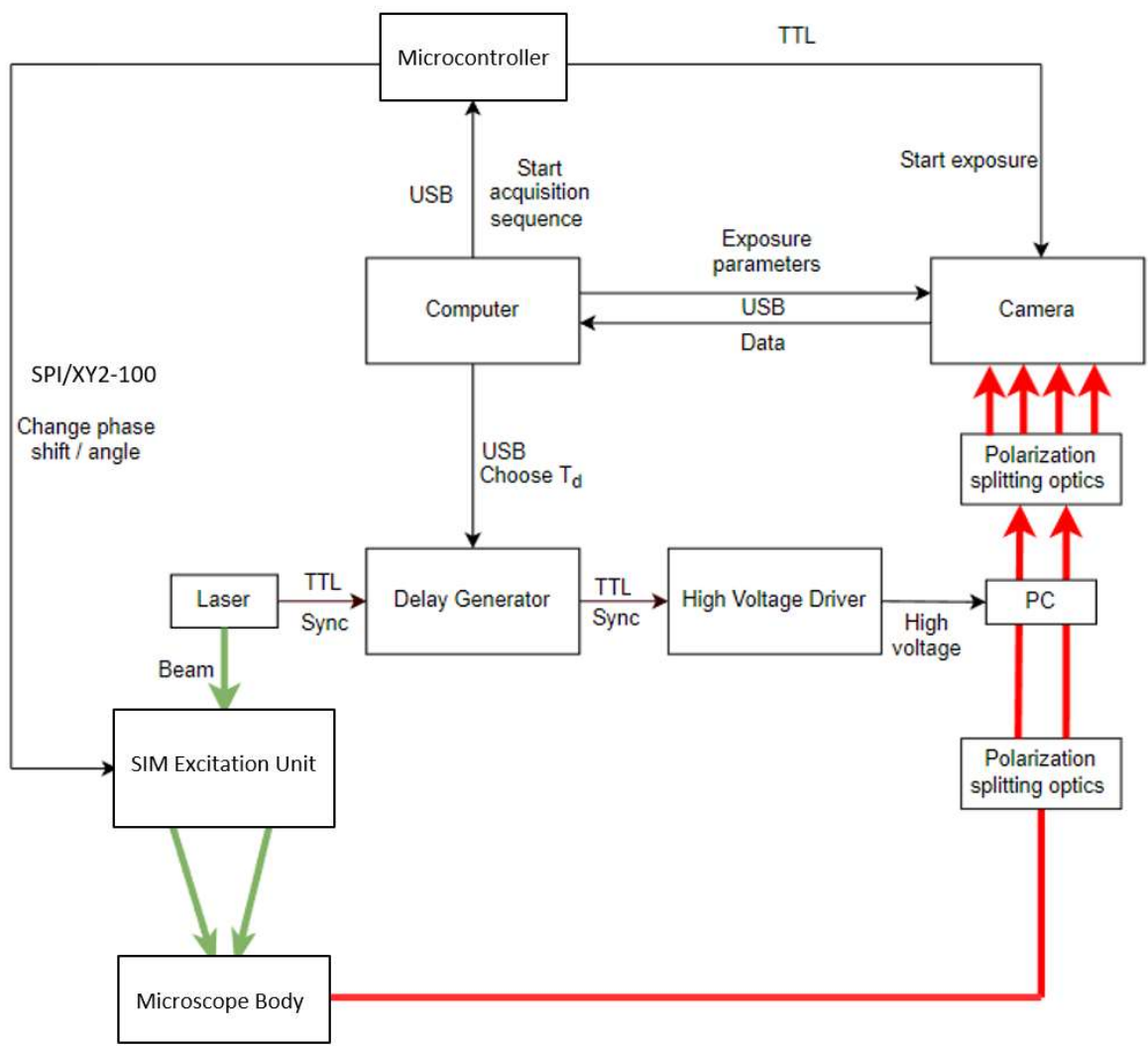

**Fig. S2: SIMFLIM experiment control.** Block diagram showing the control scheme of the setup. **FLIM:** a computer runs a python script setting the gating time of the Pockels cell via the time delay $\Delta t$ applied by the delay generator. **SIM:** Once the gating time is set, the SIM acquisition procedure starts, allowing to acquire 15 raw SIM frames, each with a specific angle (3 different angles in total) and phase (5 different phases in total) of the SIM excitation interference pattern: the computer runs a Python program setting a microcontroller responsible for the SIM raw frames acquisition. The microcontroller controls first the galvanometric mirrors of the excitation path via a XY2-100 interface, and the phase modulator via an SPI interface. The camera is triggered subsequently for the exposure to start. The camera integrates the images over multiple pulses. Then, the next raw frame is acquired, until the acquisition of the 15 raw frames is complete. The camera parameters are set, and the images are read out by MicroManager[5,6].

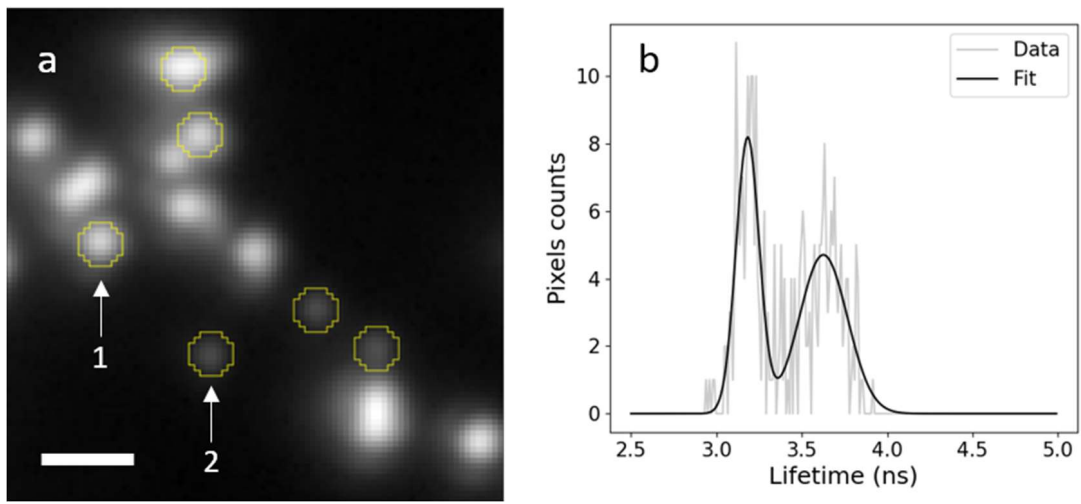

**Fig. S3: Microsphere lifetime analysis. a.** WF-FLIM Intensity channel corresponding to Fig. 2a (main text), showing TS and FS microspheres. The gamma was adjusted to 0.5 for better visualization of the TS microspheres. Scale bar: 1 μm. The maximum number of counts/pixel/s stated in the main text is measured for the single FS (resp. TS) microsphere indicated as 1 (resp. 2). **b.** Histogram of all measured lifetime values within the regions of interest (yellow circles) defined in (a). The histogram is fitted using the sum of two Gaussian functions, yielding the lifetimes $\tau_1 = 3.18\ ns$, $\tau_2 = 3.63\ ns$, and respective widths $\sigma_1 = 0.07\ ns$, $\sigma_2 = 0.14\ ns$.

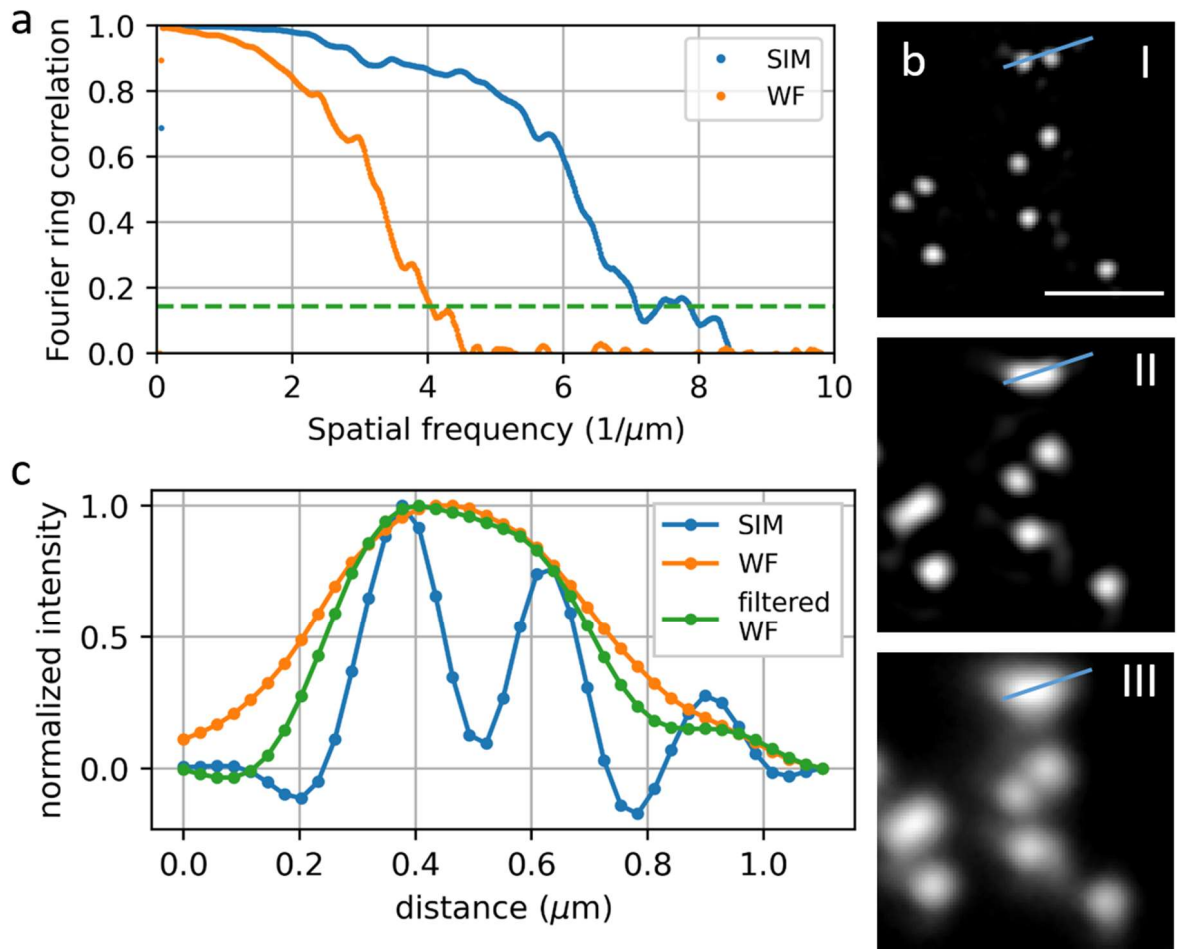

**Fig. S4: Spatial resolution analysis of the intensity channel of a SIMFLIM image. a.** Fourier ring correlation analysis[7] on the intensity channel of the SIMFLIM image of the full FOV (diameter 55 µm) with fluorescent beads corresponding to Fig. 2a. The green dashed line marks the threshold corresponding to the spatial frequency cutoff. A spatial resolution of 141 nm for SIM is observed as compared to 245 nm for WF. **b.** Intensity channel of a region of interest from Fig. 2, as observed in SIMFLIM (I), filtered WF-FLIM (II), and WF-FLIM (III). The filtered WF image is obtained by the SIM reconstruction processed in the ImageJ plugin fairSIM[8]. Scale bar: 1 µm. The best spatial resolution is observed in the SIMFLIM image. This is also observed in **c**, which shows intensity plots along the line marked in (b). While WF and filtered widefield images do not allow to discern individual microspheres which are ~200 nm apart, SIM provides the spatial resolution to do so.

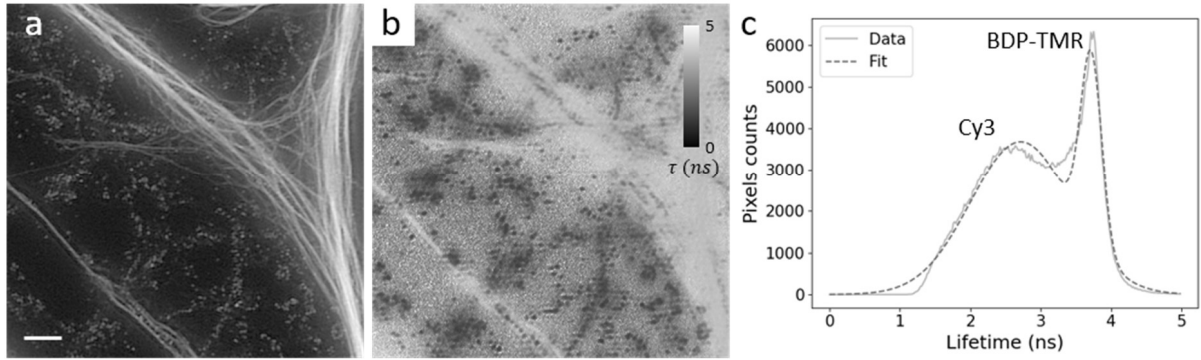

**Fig. S5: SIMFLIM lifetime analysis of the neuron culture sample discussed in Fig. 3 in the main text.**
**a.** Intensity channel of a ROI of the SIMFLIM image (gamma correction: 0.3). Scale bar: 2 μm. **b.** Corresponding lifetime channel. **c** Lifetime histogram corresponding to **b**. The experimental histogram is fitted using a sum of two Gaussian functions, giving $\tau_1 = 2.7\ ns$, $\tau_2 = 3.7\ ns$, and the respective widths $\sigma_1 = 0.7\ ns$, $\sigma_2 = 0.15\ ns$. The peak at 2.7 ns is attributed to the fluorophore Cy3 and the peak at 3.7 ns is attributed to BDP-TMR.

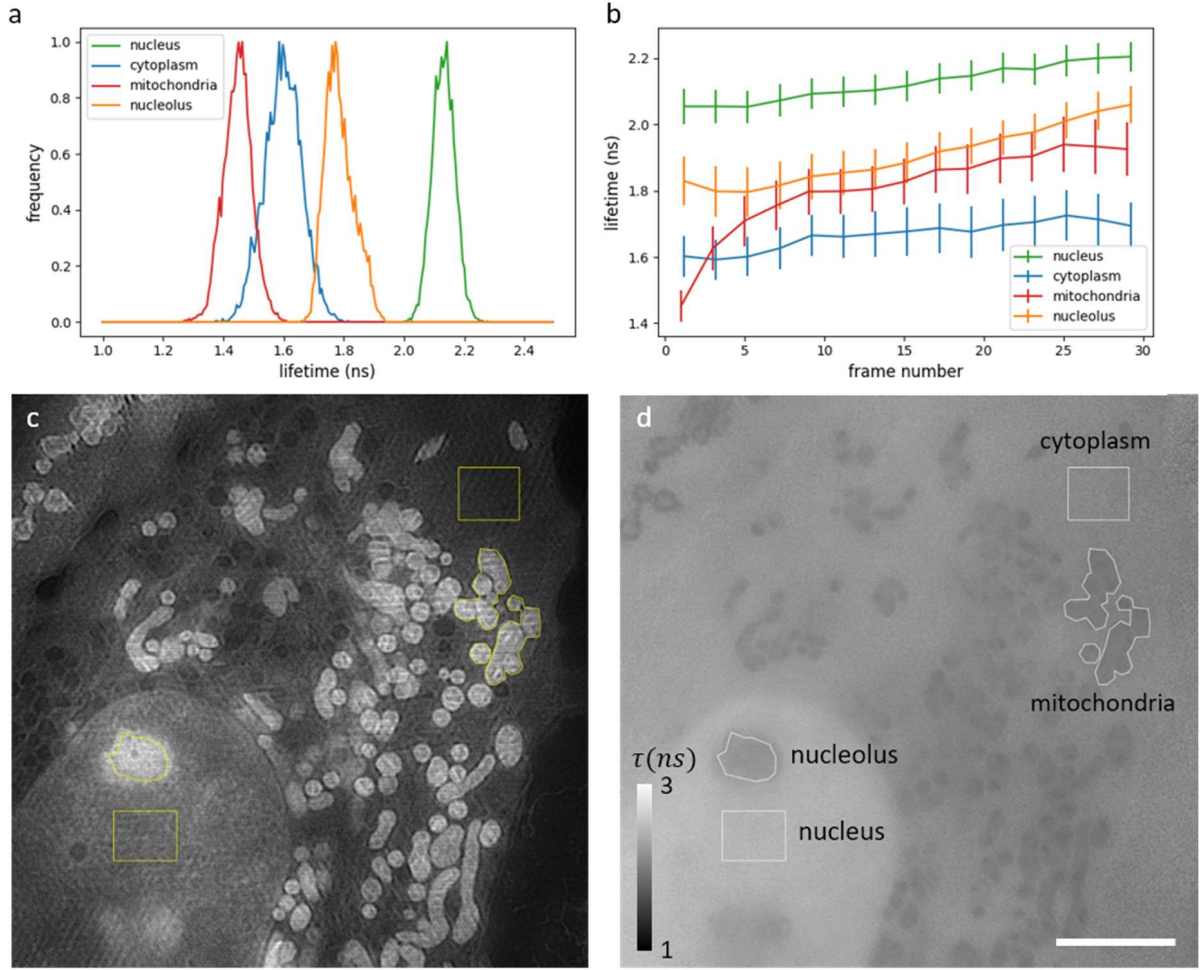

**Fig. S6: SIMFLIM lifetime analysis of the U2OS cell stained with MTO, discussed in Fig. 5 in the main text. a.** Lifetime histogram corresponding to the ROIs shown in c and d (first frame of the image sequence). Note that the absolute lifetime values also vary by +/- 10 % depending on the position on the FOV due to a systematic error (see Fig. S10) that needs to be corrected in future work. **b.** time evolution: the average lifetime of each ROI is plotted as a function of the frame number, while the ROI is updated to follow the dynamics of the cell. The error bars represent the standard deviation of the measured lifetime over the ROI. Note that the curves have been slightly shifted with respect to each other in the direction of the x axis for better visibility of the error bars. We observe an increase of lifetime in the mitochondria of ~300ps/min in the first minute of the measurement. This is significantly larger than typical thermally induced drifts observed in our instrument (~10 ps/min). **c.** Intensity channel of the first SIMFLIM frame of the image sequence corresponding to Fig. 5 in the main text, showing the ROIs (yellow) used for the histogram in a. **d.** corresponding lifetime channel, with the ROIs appearing in light gray. Scale bar: 5 µm.

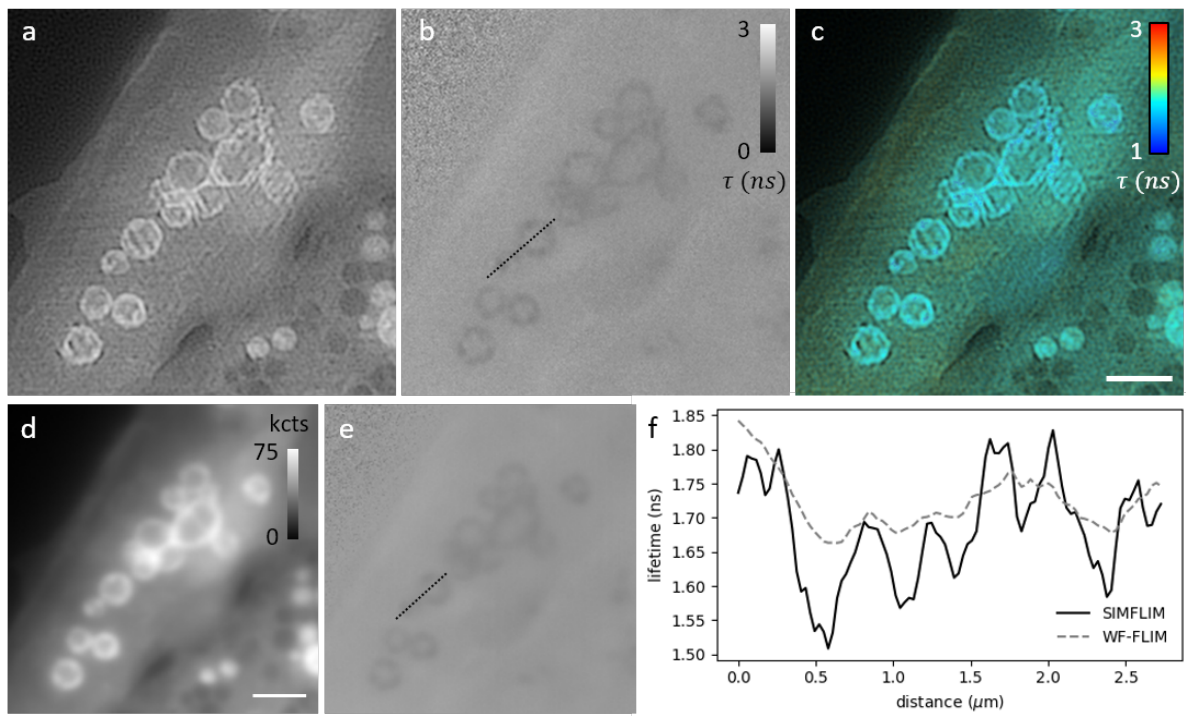

**Fig. S7: Zoomed view of round, hollow-looking mitochondria in a U2OS cell stained with MTO, discussed in Fig. 5 in the main text, and comparison of the spatial resolution in the lifetime channel of the SIMFLIM and WF-FLIM images: a., b., c.** SIMFLIM image corresponding to a ROI in the second frame of SM3. **a.** intensity channel (with gamma = 0.5). **b.** lifetime channel. **c.** corresponding composite SIMFLIM image (intensity channel with gamma = 1). Scale bar: 2 µm. **d.** intensity channel (gamma = 1) of the corresponding WF-FLIM image. Scale bar: 2 µm. The calibration bar shows the number of photons per pixel in kcounts (assuming the same pixel size as the super-resolved image). **e.** corresponding lifetime channel. **f.** cross-section along the dashed line shown in b. and e. The resolution in the lifetime channel of the full-FOV image was estimated to be 156 nm by FRC.

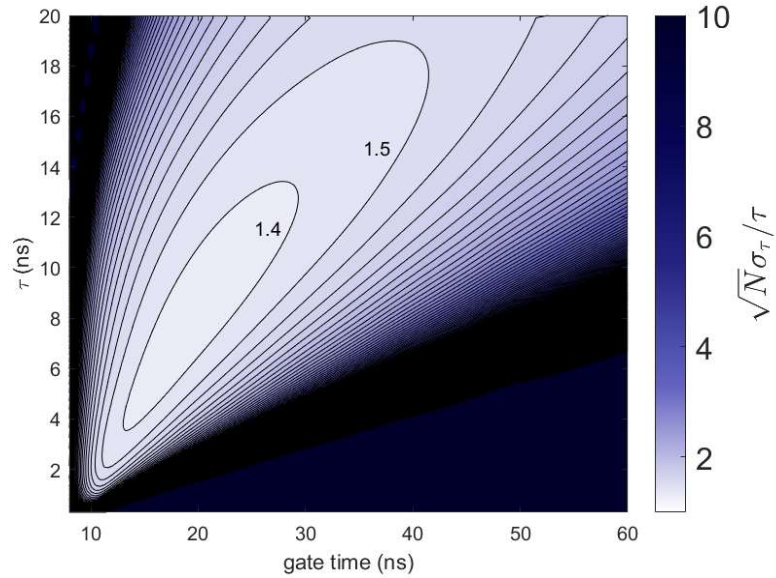

**Fig. S8: Simulated F-number (see discussion below) as a function of the experimental lifetime $\tau$ to be measured and the gating time $t_g$ (in the case when a single value of $t_g$ is used, see methods).** The 2 labels on the plot display the value of the iso-F lines. The plot was generated using the experimentally measured IRF (average ratio over a square of 670 x 670 pixels centered in the FOV). Shot noise in the gated and ungated image intensities is used to calculate uncertainty in the image intensity ratio, which is then propagated to an uncertainty in the fluorescence lifetime estimate using the slope of the ratio-to-lifetime lookup table.

The precision that can be achieved in lifetime measurements is bounded according to[9]:

$$\frac{\sigma_\tau}{\tau} = \frac{F}{\sqrt{N}}$$

where N is the number of collected photons, $\sigma_\tau$ is the standard deviation on the measured lifetime $\tau$, and F is a figure of merit dependent on the technology used, with F ≥ 1. It has been shown that 2-bin EOFLIM can reach this limit up to a factor of F = 1.3[10]. Fig. S8 shows the simulated F as a function of $\tau$ and $t_g$. In our case, the theoretical value is F=1.6 (at $t_g = 10.4\ ns, \tau = 1.8\ ns$). With N = 30,000 (for example the cytoplasm next to mitochondria in Fig. S8 c.), and $\tau = 1.8\ ns$, this yields a shot-noise limited precision of $\sigma_\tau = 17\ ps$.

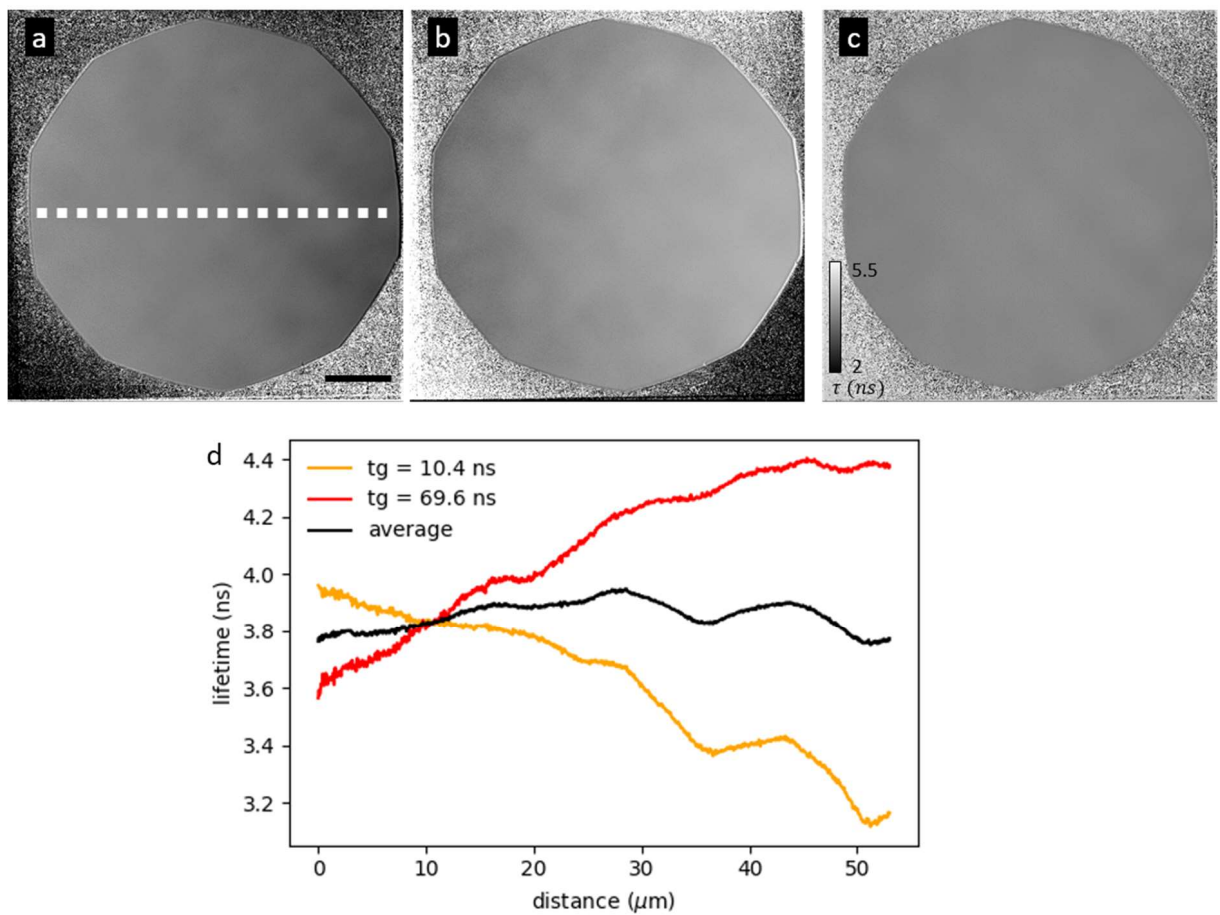

**Fig. S9: Reduction of systematic errors by combining lifetime images taken with two different gating times.** Lifetime channel of a WF-FLIM image of a solution of CellMask™ Orange Actin Tracking Stain, **a.** taken with $t_g = 10.4\ ns$, Scale bar: 10 μm, **b.** taken with $t_g = 69.6\ ns$, **c.** obtained by averaging a. and b. **d.** cross-section, along the white dashed line shown in a. (width: 50 pixels), of a, b, and c. Combining data taken with these two gating times (corresponding to the rising and the falling edge of the voltage pulse applied to the Pockels cell) allows to reduce the linear variation observed from the left to the right of the FOV. For obtaining this data, the commercial solution was first diluted 10 times in bi-distilled water. 5 μL of this solution was introduced between a microscopy glass slide and a cover glass (18 mm X 18 mm) and sealed with nail polish.

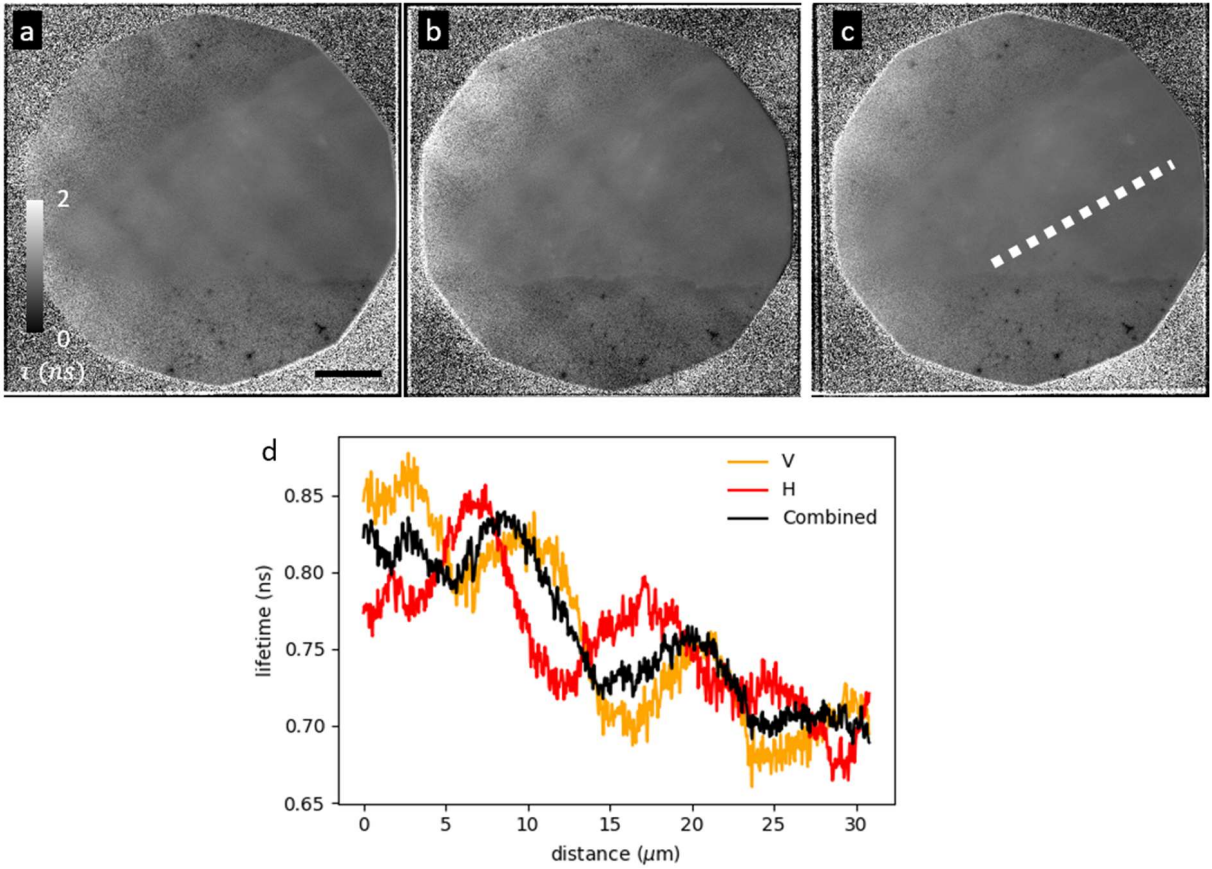

**Fig. S10: Reduction of systematic errors by combining the vertical and horizontal polarization channels.** Lifetime channels of SIMFLIM images of the first frame of the sequence of Fig. 4, obtained with the V (**a.**), H (**b.**), and combined (**c.**) polarization channel (see below). Scale bar: 10 µm. **d.** cross section along the white dashed line (width: 50 pixels) shown in c., for the V, H, and combined polarization channels. Combining the H and V polarization channels reduces the ringing contribution to systematic errors from 30 ps to 20 ps rms after subtracting a linear fit from the curves. These inhomogeneities, as well as the linear variation across the FOV (see Fig. S9), are likely due to thermal effects in the PC and the PC driver.

For a., the ratio (see data processing in methods) is defined as:

$$r_V(t_g) = \frac{I_{V,gated}(t_g)}{I_{V,gated}(t_g) + I_{V,ungated}(t_g)} = \frac{I_{VH}(t_g)}{I_{VH}(t_g) + I_{VV}(t_g)}$$

Similarly, for b., the ratio is defined as:

$$r_H(t_g) = \frac{I_{H,gated}(t_g)}{I_{H,gated}(t_g) + I_{H,ungated}(t_g)} = \frac{I_{HV}(t_g)}{I_{HV}(t_g) + I_{HH}(t_g)}$$